\documentclass[twocolumn,superscriptaddress,nolongbibliography,amsmath,amssymb,aps,prb]{revtex4-2}
\usepackage[colorlinks=true,urlcolor=blue,citecolor=blue,linkcolor=blue]{hyperref}
\usepackage{txfonts}
\usepackage{graphicx}
\usepackage{bm}
\usepackage{booktabs}
\usepackage{amsmath}
\usepackage{pgf}
\usepackage{subfigure}
\usepackage{braket}
\usepackage[normalem]{ulem}
\usepackage[mathscr]{eucal}
\usepackage{simpler-wick}
\usepackage{tikz}
\usetikzlibrary{shapes.geometric, arrows}
\tikzstyle{startstop} = [rectangle, rounded corners, minimum width=1cm, minimum height=0.3cm,text centered, draw=black, fill=red!30]
\tikzstyle{process} = [rectangle,rounded corners, minimum width=1cm, minimum height=0.3cm, text centered, draw=black, fill=orange!30]
\tikzstyle{process_same} = [rectangle,rounded corners, minimum width=6cm, minimum height=0.3cm, text centered, draw=black, fill=orange!30]
\tikzstyle{decision} = [trapezium, trapezium left angle=70, trapezium right angle=110, minimum width=1cm, minimum height=0.3cm, text centered, draw=black, fill=blue!30]
\tikzstyle{arrow} = [thick,->,>=stealth]
\usepackage{dcolumn}
\renewcommand{\vec}[1]{\mathbf{#1}}

\newcommand{\Eq}[1]{Eq.~(\ref{#1})}

\usepackage{algpseudocode}
\usepackage{algorithm}

\algnewcommand\algorithmicforeach{\textbf{for each}}
\algdef{S}[FOR]{ForEach}[1]{\algorithmicforeach\ #1\ \algorithmicdo}

\begin{document}
\title{Projected d-wave superconducting state: a fermionic projected entangled pair state study}
\author{Qi Yang}
\affiliation{Institute of Physics, Chinese Academy of Sciences, Beijing 100190, China}
\affiliation{University of Chinese Academy of Sciences, Beijing 100049, China}
\author{Xing-Yu Zhang}
\affiliation{Institute of Physics, Chinese Academy of Sciences, Beijing 100190, China}
\affiliation{University of Chinese Academy of Sciences, Beijing 100049, China}
\author{Hai-Jun Liao}
\email{navyphysics@iphy.ac.cn}
\affiliation{Institute of Physics, Chinese Academy of Sciences, Beijing 100190, China}
\affiliation{Songshan Lake Materials Laboratory, Dongguan, Guangdong 523808, China}
\author{Hong-Hao Tu}
\email{hong-hao.tu@tu-dresden.de}
\affiliation{Institute of Theoretical Physics, Technische Universit\"at Dresden, 01062 Dresden, Germany}
\author{Lei Wang}
\email{wanglei@iphy.ac.cn}
\affiliation{Institute of Physics, Chinese Academy of Sciences, Beijing 100190, China}
\affiliation{Songshan Lake Materials Laboratory, Dongguan, Guangdong 523808, China}
\date{\today}
\begin{abstract}
We investigate the physics of projected d-wave pairing states using their fermionic projected entangled pair state (fPEPS) representation. First, we approximate a d-wave Bardeen-Cooper-Schrieffer state using the Gaussian fPEPS. Next, we translate the resulting state into fPEPS tensors and implement the Gutzwiller projection which removes double occupancy by modifying the local tensor elements. The tensor network representation of the projected d-wave pairing state allows us to evaluate physical quantities in the thermodynamic limit without employing the Gutzwiller approximation. Despite having very few variational parameters, such physically motivated tensor network states are shown to exhibit competitive energies for the doped $t$-$J$ model. We expect that such construction offers useful initial states and guidance for variational tensor network calculations.
\end{abstract}
\maketitle

\section{Introduction}

The projected Bardeen-Cooper-Schrieffer (BCS) states play a prominent role in the studies of strongly correlated electrons~\cite{Lee2008}. The state is obtained by eliminating the double occupancies in the BCS wave function
\begin{equation}
|\Psi \rangle = P_G |\mathrm{BCS} \rangle,
\label{eq:p-BCS}
\end{equation}
where $P_G=\prod_{\mathbf{i}}\left(1-n_{\mathbf{i}\uparrow}n_{\mathbf{i}\downarrow}\right)$ is the Gutzwiller projection operator which implements the projection.
The state consists of resonating valence bonds (RVBs), which are believed to be relevant to superconducting cuprates~\cite{P.W.Anderson1986,Baskaran1988,Anderson2004}, frustrated magnets~\cite{Fazekas1974}, and a broad range of other phenomena in strongly correlated physics. Subsequent theoretical and numerical investigations show that the projected BCS state in \Eq{eq:p-BCS} is indeed the low-energy candidate state of the relevant $t$-$J$ models and exhibits similar features as observed in superconducting cuprates~\cite{fczhang1988, Gros1988, Gros1989, Yokoyama1988, Paramekanti2001, Edegger2007}.

To account for further intricacies such as competing orders, the vanilla RVB state~\cite{Anderson2004} has developed into a full-fledged variational wave function~\cite{Yokoyama, Ivanov2004, Lugas2006, Darmawan2018a, Misawa2019}. Along with developments of other numerical methods~\cite{Corboz2014, Zheng2017b, Ponsioen2019a, Jiang2021, Gong2021a, Jiang2021a, Qin2020b, Sorella2021}, a point has been reached where uncertainties in the model Hamiltonian are even larger than the achieved accuracy of a many-body solver. The situation calls for more realistic models with inputs from ab initio calculations and experiments. In the meantime, it is worth pursuing deeper synergy between different methods beyond simply cross-checking their numerical data~\cite{Zheng2017b}. Such synergy will bridge the worlds of ``educated guess of wave functions'' and ``solving Hamiltonians numerically'', potentially offering more physical understandings.

An example of such promising synergy is to cast the family of projected BCS states into tensor network states. Recently, a number of works have developed methods for converting projected fermionic states into matrix product states~\cite{Silvi2013a, zaletel2014exact, Fishman2015a, Bauer2019a, Wu2020g, Jin2020a, Petrica.2021, Jin2020b, Aghaei2020, Jin2021d} and used such translation to inspect state fidelity, study their entanglement properties, and facilitate density-matrix renormalization group calculations.

Extending this progress to convert projected fermionic states into two-dimensional tensor networks is highly desirable as this will allow one to investigate projected BCS states using methods from the tensor network toolbox. Along this line, there have been works constructing tensor network representations of the RVB states based on their real space picture~\cite{Verstraete2006a,Poilblanc2012, Poilblanc2014}. In this work, we present a more generic approach which is based on Gaussian fermionic projected entangled pair state (fPEPS)~\cite{Kraus2010} and has a variable number of parameters to achieve the translation. Then we use it to investigate the projected d-wave pairing state in an infinite two-dimensional lattice.
\tikzset{
master/.style={
execute at end picture={
\coordinate (lower right) at (current bounding box.south east);
\coordinate (upper left) at (current bounding box.north west);
\coordinate (lower left) at (current bounding box.south west);
\coordinate (upper right) at (current bounding box.north east);
}
},
slave/.style={
execute at end picture={
\pgfresetboundingbox

\path (upper right) rectangle (lower left);
}
}
}
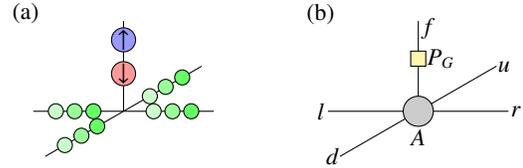
\begin{figure}[t]
{
\begin{tikzpicture}[master]
\draw (0,0) -- (0.3*4,0);
\draw (0,0) -- (-0.3*4,0);
\draw (0,0) -- (0.260*4,0.15*4);
\draw (0,0) -- (-0.260*4,-0.15*4);
\draw (0,0) -- (0,1.2);
\filldraw[fill=green!20!white, draw=black] (0.4,0) circle (0.1cm);
\filldraw[fill=green!40!white, draw=black] (0.65,0) circle (0.1cm);
\filldraw[fill=green!60!white, draw=black] (0.9,0) circle (0.1cm);
\filldraw[fill=green!60!white, draw=black] (-0.4,0) circle (0.1cm);
\filldraw[fill=green!40!white, draw=black] (-0.65,0) circle (0.1cm);
\filldraw[fill=green!20!white, draw=black] (-0.9,0) circle (0.1cm);
\filldraw[fill=green!60!white, draw=black] (-0.35,-0.2) circle (0.1cm);
\filldraw[fill=green!40!white, draw=black] (-0.56,-0.325) circle (0.1cm);
\filldraw[fill=green!20!white, draw=black] (-0.78,-0.45) circle (0.1cm);
\filldraw[fill=green!20!white, draw=black] (0.35,0.2) circle (0.1cm);
\filldraw[fill=green!40!white, draw=black] (0.56,0.325) circle (0.1cm);
\filldraw[fill=green!60!white, draw=black] (0.78,0.45) circle (0.1cm);
\filldraw[fill=red!40!white, draw=black] (0,0.5) circle (0.16cm);
\node at (0,0.5) {$\downarrow$};
\filldraw[fill=blue!40!white, draw=black] (0,0.95) circle (0.16cm);
\node at (0,0.95) {$\uparrow$};
\node at (-1.3,1.3) {(a)};
\filldraw[fill=none, draw=none] (2,0) circle (0.1cm);
\end{tikzpicture}
}
{
\begin{tikzpicture}[slave]
\draw (0.1,0) -- (0.3*4,0);
\draw (-0.1,0) -- (-0.3*4,0);
\draw (0.0866,0.05) -- (0.260*4,0.15*4);
\draw (-0.0866,-0.05) -- (-0.260*4,-0.15*4);
\draw (0,0.1) -- (0,0.6);
\filldraw[fill=yellow!40!white, draw=black] (-0.10,0.8) rectangle (0.1,0.6);
\draw (0,0.8) -- (0,1.2);
\node at (0.3,0.7) {$P_G$};
\filldraw[fill=gray!40!white, draw=black] (0,0) circle (0.2cm);
\node at (0,-0.35) {$A$};
\node at (0.3*4+0.1,0) {$r$};
\node at (-0.3*4-0.1,0) {$l$};
\node at (0.260*4+0.1,0.15*4) {$u$};
\node at (-0.260*4-0.1,-0.15*4) {$d$};
\node at (0.13,1.1) {$f$};
\node at (-1.3,1.3) {(b)};
\end{tikzpicture}
}
\caption{(a) The local fiducial state in \Eq{eq:LocalState} of a fPEPS. It is a state formed by the physical fermions of spin up and down (blue and red dots) and $\Lambda$ flavors of virtual fermions reside on the bond (green dots). The state is fully characterized by the local correlation matrix in \Eq{eq:LocalGamma} if it is Gaussian. (b) Gutzwiller projection to the local fPEPS tensor.} \label{fig:fpeps-Gutzwiller}
\end{figure}

Our starting point is a fermionic quadratic Hamiltonian with d-wave pairing on the two-dimensional (2D) square lattice,
\begin{equation}
H_\mathrm{BCS}=\sum_{\mathbf{k},\sigma}  \xi_{\mathbf{k}} f_{\mathbf{k} \sigma}^{\dagger} f_{\mathbf{k} \sigma}
+ \sum_{\mathbf{k}} \left(\Delta_{\mathbf{k}} f_{\mathbf{k} \uparrow}^{\dagger} f_{-\mathbf{k} \downarrow}^{\dagger}+\Delta_{\mathbf{k}}^{*} f_{-\mathbf{k} \downarrow} f_{\mathbf{k}\uparrow} \right),
\label{eq:HMF}
\end{equation}
where $f_{\mathbf{k}\sigma}$ ($f^{\dag}_{\mathbf{k}\sigma}$) is the fermion annihilation (creation) operator in momentum space with spin index $\sigma=\uparrow,\downarrow$, $\xi_{\mathbf{k}}=-2t(\cos{k_x}+\cos{k_y})-\mu$ and $\Delta_{\mathbf{k}}=2\Delta_d (\cos{k_x}-\cos{k_y})$. The hopping amplitude $t$ is set to unity. The state $\ket{\mathrm{BCS}}$ is the ground state of $H_{\mathrm{BCS}}$ parametrized by the chemical potential $\mu$ and the d-wave pairing amplitude $\Delta_d$.

Given the BCS Hamiltonian in \Eq{eq:HMF}, we first find a Gaussian fPEPS approximation to the pairing state $\ket{\mathrm{BCS}}$ via a variational calculation. As shown in Fig.~\ref{fig:fpeps-Gutzwiller}(a), the Gaussian fPEPS is formed by the physical fermions of spin up and down (blue and red dots) and $\Lambda$ flavors of virtual fermions residing on each bond (green dots). Next, we translate the Gaussian fPEPS, parametrized by its correlation matrix, to fPEPS tensors and implement the Gutzwiller projection by locally modifying the tensor elements. Having an approximated fPEPS representation of the projected BCS state in \Eq{eq:p-BCS} allows us to investigate its properties using tensor network algorithms. We examine the validity of the Gutzwiller approximation for the hole fugacity and report on the pairing order parameter and variational energy of the projected d-wave pairing state for the $t$-$J$ model with various dopings.
\section{Method\label{sec:method}}

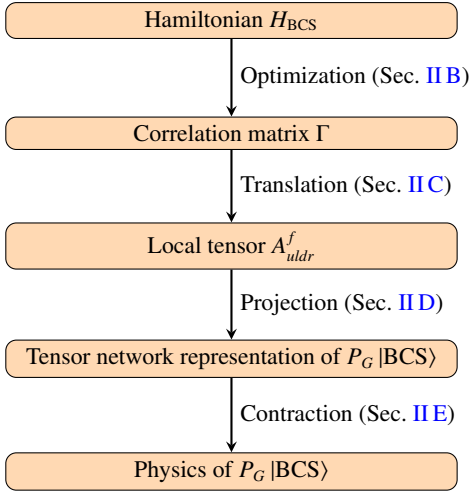
\begin{figure}
  \begin{tikzpicture}[node distance=1.5cm]
      \node (start) [process_same] {Hamiltonian $H_\mathrm{BCS}$ };
      \node (pro1) [process_same, below of=start] {Correlation matrix $\Gamma$};
      \node (pro2) [process_same, below of=pro1] {Local tensor $A^f_{uldr}$};
      \node (dec1) [process_same, below of=pro2] {Tensor network representation of $P_G\ket{\mathrm{BCS}}$};
      \node (phy) [process_same, below of=dec1] {Physics of $P_G\ket{\mathrm{BCS}}$};

      \draw [arrow] (start) -- node[anchor=west] {Optimization (Sec.~\ref{sec:ABD})} (pro1);
      \draw [arrow] (pro1) -- node[anchor=west] {Translation (Sec.~\ref{sec:translation})} (pro2);
      \draw [arrow] (pro2) -- node[anchor=west] {Projection (Sec.~\ref{sec:projection})} (dec1);
      \draw [arrow] (dec1) -- node[anchor=west] {Contraction (Sec.~\ref{sec:contraction})} (phy);
  \end{tikzpicture}
  \vspace{4mm}
  \caption{The recipe for preparing a Gutzwiller projected BCS state in \Eq{eq:p-BCS} as a fermionic PEPS and investigating its physical properties via tensor network contraction.}
  \label{fig:workflow}
\end{figure}

In this section, we present our method for obtaining the fPEPS tensors of a projected BCS state and the workflow is shown in Fig.~\ref{fig:workflow}. This section is organized as follows: We first review the construction of Gaussian fPEPS~\cite{Kraus2010} in Sec.~\ref{sec:construction_of_gfpeps} and the variational optimization of the Gaussian fPEPS~\cite{Mortier2020} in Sec.~\ref{sec:ABD}. Then we develop the method to cast the Gaussian fPEPS to fPEPS tensors and implement the Gutzwiller projection in Sec.~\ref{sec:translation} and Sec.~\ref{sec:projection}, respectively. Finally, in Sec.~\ref{sec:contraction}, we discuss how to make the fPEPS tensor obtained in Sec.~\ref{sec:projection} compatible with the method of contracting fPEPS tensors developed in Ref.~\cite{Corboz2010}.

\subsection{Construction of Gaussian fPEPS\label{sec:construction_of_gfpeps}}

The fPEPS is formed by mediating the entanglement of physical fermions ($f$) on lattice sites via virtual fermions $(u,l,d,r)$ residing on the bonds. Formally, it can be expressed as~\cite{Kraus2010,Wahl2014}
\begin{equation}
\ket{\Psi}=\left[\bigotimes_{\left\langle\mathbf{i}, \mathbf{j} \right\rangle}\bra{\omega_{\mathbf{i} \mathbf{j} }} \right] \left[ \bigotimes_{\mathbf{i}} \ket{A_{\mathbf{i}} } \right],
\label{eq:fpeps}
\end{equation}
where $|\omega_{\mathbf{i} \mathbf{j}}\rangle$ is a maximally entangled state of virtual fermions on the $\langle \mathbf{i},  \mathbf{j} \rangle$ bond. More concretely, we use $u_{\mathbf{i} \alpha}, l_{\mathbf{i} \alpha}, d_{\mathbf{i} \alpha}, r_{\mathbf{i} \alpha}$ to denote the annihilation operators of the virtual fermions at site $\mathbf{i}$, where $\alpha \in [1, \ldots , \Lambda ]$ is the flavor index and $u,l,d,r$ refer to up, left, down, and right, respectively. The horizontal bond between two neighboring sites $\mathbf{i}$ and $\mathbf{j} = \mathbf{i} + \hat{\mathbf{x}}$ is defined by
\begin{equation}
\ket{\omega_{\mathbf{i} \mathbf{j} }}=\prod_{\alpha=1}^{\Lambda} \frac{1}{\sqrt{2}}\left(1+ r_{ \mathbf{i} \alpha }^{\dagger}l_{\mathbf{j} \alpha }^{\dagger}\right)| \mathrm{vac} \rangle,
\label{eq:omegaij}
\end{equation}
where $\ket{ \mathrm{vac} }$ is the vacuum of virtual fermions. The vertical bonds are defined in a similar way.

Next, the local fiducial state~\cite{Wahl2014} is generally expressed as
\begin{equation}
\label{eq:LocalState}
\ket{A_{\mathbf{i}}} = \sum_{f,u,l,d,r} A^{f}_{uldr} \ket{f}\otimes \ket{uldr},
\end{equation}
where $\ket{f}$ denotes the local physical states, consisting of empty, singly occupied (with either spin up or down), and doubly occupied states. The basis $\ket{uldr}$ corresponds to the Fock space of virtual fermions at site $\mathbf{i}$.

Since the virtual bonds are fermionic Gaussian states, the fPEPS in Eq.~\eqref{eq:fpeps} would also be Gaussian, as long as the local fiducial state $\ket{A_{\mathbf{i}}}$ in \Eq{eq:LocalState} is a fermionic Gaussian state. Such a state forms a subclass of fPEPS, namely the Gaussian fPEPS~\cite{Kraus2010}. The flavor number of virtual fermions controls the expressibility of the Gaussian fPEPS. In the usual fPEPS language, a Gaussian fPEPS with flavor number $\Lambda$ has a bond dimension $D=2^{\Lambda}$.

A great computational advantage of the Gaussian fPEPS is that a powerful fermionic Gaussian formalism~\cite{Bravyi2005} can be used for performing efficient computations, which we briefly outline below. Throughout this work we consider translationally invariant fPEPS on the 2D square lattice with a one-site unit cell (generalization to multi-site unit cells and/or other lattices is straightforward though). Let us omit the site index for now and label the physical and virtual fermions at a local site as $(c_1,c_2,c_3,...,c_{4\Lambda+2})=(f_\uparrow,f_\downarrow, r_{1},l_{1},r_{2},l_{2},...,r_{\Lambda},l_{\Lambda},
d_{1},u_{1},d_{2},u_{2},...,d_{\Lambda},u_{\Lambda})$. In terms of Majorana operators $\gamma_{2\mu-1} = c_\mu^\dagger+c_\mu$ and $\gamma_{2\mu} =-i(c_\mu^\dagger-c_\mu)$, the local fiducial state~\eqref{eq:LocalState}, being a fermionic Gaussian state, is fully characterized by its correlation matrix~\cite{Bravyi2005}
\begin{equation}
\label{eq:LocalGamma}
\Gamma_{\mu \nu} = \frac{i}{2} \langle A | [\gamma_\mu,\gamma_\nu] | A \rangle,
\end{equation}
where $\Gamma$ is a real antisymmetric matrix satisfying $\Gamma^2 = -\mathbb{I}_{(8\Lambda+4)\times (8\Lambda+4)}$. As such, the correlation matrix~\eqref{eq:LocalGamma} can be written as
\begin{equation}
\label{eq:Gamma-ABD}
\Gamma = \begin{pmatrix} A & B \\ -B^T & D \end{pmatrix},
\end{equation}
where $A$ and $D$ are $4\times 4$ and $8\Lambda\times 8\Lambda$ real antisymmetric matrices, respectively, and $B$ is a $4\times 8\Lambda$ real matrix.

In a translationally invariant setup, both physical and virtual Majorana modes can be transformed into modes in momentum space with $\gamma_{\vec{k},\mu}=\frac{1}{\sqrt{N}}\sum_{\vec{j}}\exp(-i\vec{k}\cdot\vec{j})\gamma_{\vec{j},\mu}$ where $N=L^2$ is the total number of sites ($L$: number of sites along $\mathbf{x}$ and $\mathbf{y}$ directions). The allowed values of $\vec{k}$ satisfy $\vec{k}=\frac{2\pi \vec{n}}{L}$ where $\vec{n}\in \mathbb{Z}\oplus (\mathbb{Z}+\frac{1}{2})$ if the system has periodic (antiperiodic) boundary condition along $\mathbf{x}$ ($\mathbf{y}$) direction~\footnote{Together with suitably chosen system sizes, this boundary condition avoids the complication of dealing with exact zero modes at the Dirac points $\mathbf{k}=(\pm \tfrac{\pi}{2},\pm \tfrac{\pi}{2})$ of the d-wave pairing Hamiltonian.}. Denoting the virtual bond state as $|\omega\rangle=\bigotimes_{\langle \mathbf{i},\mathbf{j}\rangle}\ket{\omega_{\mathbf{ij}}}$, the correlation matrix of the virtual bond state is written as $(G^{\omega}_{\vec{k}})_{\mu\nu}= \frac{i}{2} \langle\omega| [\gamma_{\vec{k},\mu},\gamma_{-\vec{k},\nu}] |\omega\rangle$ ($\mu,\nu = 5,\ldots,8\Lambda+4$) and takes the following explicit form~\cite{Mortier2020}:
\begin{equation}
G^{\omega}_{\vec{k}}=\left[\bigoplus_{\mu=1}^{\Lambda}
\begin{pmatrix} & e^{ik_x}\sigma^x \\
-e^{-ik_x}\sigma^x &  \\
\end{pmatrix}\right]\oplus\left[\bigoplus_{\mu=1}^{\Lambda}
\begin{pmatrix} & e^{ik_y}\sigma^x \\
-e^{-ik_y}\sigma^x   & \\
\end{pmatrix}\right]
\end{equation}
where $\sigma^x$ is the Pauli matrix.

The correlation matrix of the Gaussian fPEPS~\eqref{eq:fpeps}, defined by
\begin{equation}
(G^{\mathrm{f}}_{\vec{k}})_{\mu\nu}= \frac{i}{2} \langle\Psi| [\gamma_{\vec{k},\mu},\gamma_{-\vec{k},\nu}] |\Psi\rangle, \; (\mu,\nu = 1,...,4),
\label{eq:Gfk}
\end{equation}
is calculated by contracting the virtual modes and gives rise to~\cite{Kraus2010}
\begin{equation}
G^{\mathrm{f}}_{\vec{k}}=A+B(D+ G^\omega_{\vec{k}})^{-1}B^T.
\label{eq:ABD}
\end{equation}
This is a key result for the variational calculation below. We provide a comprehensive proof of Eq.~\eqref{eq:ABD} in Appendix \ref{appendix:proof-abd}.

\subsection{Variational optimization of Gaussian fPEPS \label{sec:ABD}}

For the next step, we determine an optimal Gaussian fiducial state~\eqref{eq:LocalState}, such that the variational energy of the Hamiltonian $H_\mathrm{BCS}$ [see Eq.~\eqref{eq:HMF}] is minimized within the family of Gaussian fPEPS for a given flavor number $\Lambda$. The BCS ground state of $H_\mathrm{BCS}$ is thus approximated by a Gaussian fPEPS. By using the correlation matrix in \Eq{eq:Gfk}, the variational energy can be expressed as
\begin{equation}
\langle H_{\mathrm{BCS}}\rangle= \sum_{\vec{k}}\xi_{\vec{k}}(\rho_{\vec{k}\uparrow}+\rho_{\vec{k}\downarrow})+\Delta_{\vec{k}}\eta_{\vec{k}}+\Delta_{\vec{k}}^{*} \eta^*_{\vec{k}}
\label{eq:expectation_HBCS}
\end{equation}
with $\rho_{\vec{k}\sigma}=\langle f_{\vec{k}\sigma}^\dagger f_{\vec{k}\sigma}\rangle$ and $\eta_{\vec{k}}=\langle f^\dagger_{\vec{k}\uparrow} f^\dagger_{-\vec{k}\downarrow}\rangle$, which can be evaluated via the relation $\sum_{\vec{k}}\xi_{\vec{k}} \langle f_{\vec{k}\uparrow}^\dagger f_{\vec{k}\uparrow} \rangle = \sum_{\vec{k}}\xi_{\vec{k}}[\frac{1}{2}-\frac{1}{2}(G_{\vec{k}}^\mathrm{f})_{1,2}]$,
$\sum_{\vec{k}}\xi_{\vec{k}} \langle f_{\vec{k}\downarrow}^\dagger f_{\vec{k}\downarrow} \rangle = \sum_{\vec{k}}\xi_{\vec{k}}[\frac{1}{2}-\frac{1}{2}(G_{\vec{k}}^\mathrm{f})_{3,4}]$ and $\eta_{\vec{k}}=\langle f^\dagger_{\vec{k}\uparrow} f^\dagger_{-\vec{k}\downarrow}\rangle=\frac{1}{4}[(G_{\vec{k}}^\mathrm{f})_{1,4}+(G_{\vec{k}}^\mathrm{f})_{2,3}+i(G_{\vec{k}}^\mathrm{f})_{2,4}-i(G_{\vec{k}}^\mathrm{f})_{1,3}]$.

We carry out the optimization following Ref.~\cite{Mortier2020}. The minimization is performed through the fiducial state's correlation matrix $\Gamma$ in Eq.~\eqref{eq:Gamma-ABD}, which can be brought into a canonical form
\begin{equation}
\Gamma=X^T\left[\bigoplus_{\eta=1}^{4\Lambda+2}\begin{pmatrix}0&1\\-1&0\end{pmatrix}\right]X,
\label{eq:XJX}
\end{equation}
where $X$ is an orthogonal matrix. The matrices $A,B,D$ in Eq.~\eqref{eq:Gamma-ABD} are hence functions of $X$, and so are the correlation matrix $G^f_{\vec{k}}$ in \Eq{eq:ABD} and the variational energy in Eq.~\eqref{eq:expectation_HBCS}.

We optimize the real orthogonal matrix $X$ by minimizing the loss function using optimization on the Stiefel manifold. The derivatives are calculated by using automatic differentiation with \texttt{JAX}~\cite{jax2018github}. The optimization algorithm is the conjugate gradient method offered by \texttt{pymanopt}~\cite{pymanopt}.
This calculation was done for a finite lattice in the momentum space. Since the computational complexity scales linearly with the system size, one can reach pretty large systems easily~\cite{Mortier2020}. The accuracy of such variational calculation can be verified by comparing the results with the exact solution of the BCS Hamiltonian.

\subsection{Translation of correlation matrix $\Gamma$ into local tensor \label{sec:translation}}

To obtain the local tensor of Gaussian fPEPS $A^f_{uldr}= \bra{f} \otimes \braket{uldr|A_{\mathbf{i}}}$ [see Eq.~\eqref{eq:LocalState}], we need to find out the fiducial state $\ket{A_{\mathbf{i}}}$ based on its correlation matrix $\Gamma$.

To this end, we simply construct a single-site fiducial Hamiltonian
\begin{equation}
h = -\sum_{\mu\nu}i\Gamma_{\mu\nu}\gamma_\mu\gamma_\nu
\label{eq:fiducial-Hamiltonian}
\end{equation}
using the correlation matrix for the fiducial state. This Hamiltonian is quadratic and contains $8\Lambda+4$ types of physical and virtual Majorana fermions at a single site.

It is easy to prove that $\ket{A_{\mathbf{i}}}$ is the unique ground state of the fiducial Hamiltonian $h$. By using the orthogonal matrix defined in \Eq{eq:XJX}, the Majorana operators can be rotated into a new basis as $\gamma'=X\gamma$. Notice that $(X\Gamma X^T)_{\mu\nu} = \frac{i}{2}\bra{A_{\mathbf{i}}}[\gamma'_\mu,\gamma'_\nu]\ket{A_{\mathbf{i}}}=\left[\bigoplus_{\eta=1}^{4\Lambda+2} \begin{pmatrix}0 & 1\\ -1 &0 \end{pmatrix}\right]_{\mu\nu}$, which shows that the Gaussian fiducial state $\ket{A_{\mathbf{i}}}$ satisfies $i\gamma'_{2m-1} \gamma'_{2m}\ket{A_{\mathbf{i}}}=\ket{A_{\mathbf{i}}}$ ($m = 1,\ldots,4\Lambda+2$). As $i\gamma'_{2m-1} \gamma'_{2m}$ is a fermion parity operator and has eigenvalues $\pm 1$, the fiducial Hamiltonian $h=-i\sum_{m}\gamma'_{2m-1}\gamma'_{2m} = -\sum_{\mu\nu}i\Gamma_{\mu\nu}\gamma_\mu\gamma_\nu$ has $\ket{A_{\mathbf{i}}}$ as its unique ground state.

We convert the fiducial Hamiltonian $h$ from the Majorana basis to the original complex fermion basis [see Eq.~\eqref{eq:LocalState}] and find its ground state by diagonalizing $h$. By reshaping the state vector $\in \mathbb{C}^{2^{4\Lambda+2}}$ into a five-leg tensor, we obtain $A^f_{uldr}$ up to an unimportant overall phase.

Since the quadratic Hamiltonian conserves the fermion parity, its ground state $\ket{A_\mathbf{i}}$ has a definite parity. Therefore, the tensor $A^{f}_{uldr}$ automatically inherits $\mathbb{Z}_2$ symmetry as the parity of the state $\ket{A_\mathbf{i}}$.

For obtaining the explicit form of $\ket{A_\mathbf{i}}$, an exact diagonalization of the fiducial Hamiltonian $h$ is managable when the number of physical and virtual modes at each site is relatively small. This is indeed the case for our benchmark example, which already shows good performance. When the number of modes at each site is large, we provide a more efficient approach based on the state overlap in Appendix~\ref{appendix:overlap-translation}.

\subsection{Gutzwiller projection of Gaussian fPEPS\label{sec:projection}}

After obtaining a Gaussian fPEPS representation of the BCS state, we implement the Gutzwiller projection on the physical leg of the local tensor $A^f_{uldr}$ as shown in Fig.~\ref{fig:fpeps-Gutzwiller}(b). Since we carry out Gutzwiller projection in an infinite system in the grand canonical ensemble, we include a fugacity term in the empty configuration to account for the change of the particle density caused by the removal of double occupancies~\cite{Bernevig2003,Edegger2005,anderson2006theory,Chou2012}. In this way, the projection operator reads
\begin{equation}
  P_G=\prod_{\mathbf{i}} z^{\left(1-n_{\mathbf{i}\uparrow}-n_{\mathbf{i}\downarrow}\right) / 2}\left(1-n_{\mathbf{i} \uparrow} n_{\mathbf{i} \downarrow}\right),
\end{equation}
where $z$ is the fugacity of holes. The Gutzwiller projection operator is a local gate acting on the physical leg with the matrix representation $\mathrm{diag}(\sqrt{z}, 1, 1, 0)$, where the basis of this representation is $(\ket{0},\ket{\uparrow},\ket{\downarrow},\ket{\uparrow\downarrow})$. The zero in the Gutzwiller projection gate sets the local tensor elements associated with doubly occupied physical fermions to zero.

In principle, in a grand canonical calculation, one needs to tune the hole fugacity to maintain the average particle number. The Gutzwiller approximation~\cite{Edegger2005} provides an estimation of the fugacity
\begin{equation}
z= \frac{2\delta}{1+\delta},
\label{eq:holefugacity}
\end{equation}
where $\delta$ is the hole density. One sees that in the undoped case, $\delta=0$, the projection removes both empty and double occupancy configurations and only keeps singly occupied sites.

\subsection{Tensor network contraction~\label{sec:contraction}}

The Gaussian fPEPS after the Gutzwiller projection is no longer a Gaussian state, but preserves the fPEPS nature. To investigate its physical properties, we employ tensor network contraction algorithms for fPEPS~\cite{Corboz2010}. Since the optimization in Sec.~\ref{sec:ABD} can be done on a sufficiently large lattice, it is essentially free of finite-size error. Thus, after translation and Gutzwiller projection, we assemble the local tensors into an infinite lattice and employ infinite tensor network contraction algorithms. Note that one needs to take care of the swap gate~\cite{Corboz.2009} when contracting the local tensors since it is a  fermionic tensor network.

To make the local tensor fully compatible with the diagrammatic notation for fPEPS in Ref.~\cite{Corboz2010}, one should pay special attention to the fermion sign associated with the order and form of virtual fermions in the definition of $\ket{A_\mathbf{i}}$ and $\ket{\omega_{\mathbf{i}\mathbf{j}}}$.

For example, we need to make sure that the order of fermionic operators in the definition of $\ket{A_\mathbf{i}}$ in Gaussian fPEPS and fPEPS are the same. In Ref.~\cite{Corboz2010}, the ordering of fermions in the local tensor is $\ket{ulfdr}$, while the tensor we obtained in Sec.~\ref{sec:translation} follows the order $(c_1,c_2,...,c_{4\Lambda+2})=(f_\uparrow,f_\downarrow, r_{1},l_{1},r_{2},l_{2},...,r_{\Lambda},l_{\Lambda},
d_{1},u_{1},d_{2},u_{2},...,d_{\Lambda},u_{\Lambda})$. In order to account for this difference, we introduce swap gates to transform between these two different orderings.

Additionally, in Ref.~\cite{Corboz2010} it is free to contract the virtual bonds of two nearby fPEPS tensors without generating swap gates. This implies that the adopted definition of virtual fermion maximally entangled state is defined differently from \Eq{eq:omegaij}. For example, for the bond between site $\mathbf{i}$ and $\mathbf{j}= \mathbf{i}+\hat{\mathbf{x}}$, one has
\begin{equation}
\ket{\omega^{\text{C}}_{\mathbf{i}\mathbf{j}}} = \frac{1}{\sqrt{2^\Lambda}}\sum_{\{ n_\alpha\}}  \prod_{\alpha=1}^{\Lambda} \left(r_{\mathbf{i}\alpha }^{\dagger}\right)^{n_\alpha} \prod_{\alpha=1}^{\Lambda} \left(l_{\mathbf{j}\alpha}^{\dagger} \right)^{n_\alpha}\ket{\mathrm{vac}},
\label{eq:omega_fpeps}
\end{equation}
where $n_\alpha\in \{0,1\}$ is the occupation number of virtual fermions.

This differs from  \Eq{eq:omegaij} by the ordering of virtual fermions if the flavor number $\Lambda > 1$. The differences between $\ket{\omega^\mathrm{C}_{\mathbf{ij}}}$ and $\ket{\omega_{\mathbf{ij}}}$ can be expressed explicitly as
\begin{equation}
  \ket{\omega^\mathrm{C}_{\mathbf{ij}}} = (-1)^{\sum_{\alpha}^\Lambda  \sum^\Lambda_{\beta<\alpha}  (r_{\mathbf{i}\alpha}^\dagger r_{\mathbf{i}\alpha})(l^\dagger_{\mathbf{j}\beta}l_{\mathbf{j}\beta})}\ket{\omega_{\mathbf{ij}}}
  \label{eq:omega_diff}
\end{equation}
for the neighboring sites $\mathbf{i}$ and $\mathbf{j} = \mathbf{i} + \hat{\mathbf{x}}$. For the case of $\mathbf{j} = \mathbf{i} + \hat{\mathbf{y}}$, $r_{\mathbf{i}\alpha}$ and $l_{\mathbf{j}\alpha}$ are replaced by $u_{\mathbf{i}\alpha}$ and $d_{\mathbf{j}\alpha}$, respectively.

All coefficients in \Eq{eq:omega_diff} can be absorbed into the local tensor $A^f_{uldr}$. Afterwards, the tensor network state is compatible with the fPEPS convention in Ref.~\cite{Corboz2010} and the resulting local tensor is suitable for an fPEPS tensor contraction code.

To compute expectation values, we consider an infinite fPEPS and adopt the variational uniform matrix product state method~\cite{vumps} to obtain the environment tensors of infinite tensor networks~\cite{Vanhecke2021TangentspaceMF}. We note that other methods (e.g., corner transfer matrix renormalization group method~\cite{nishino1997corner,orus2009}) are also applicable.
\section{Results}
\begin{figure}[t]
\centering

\includegraphics[width=\columnwidth]{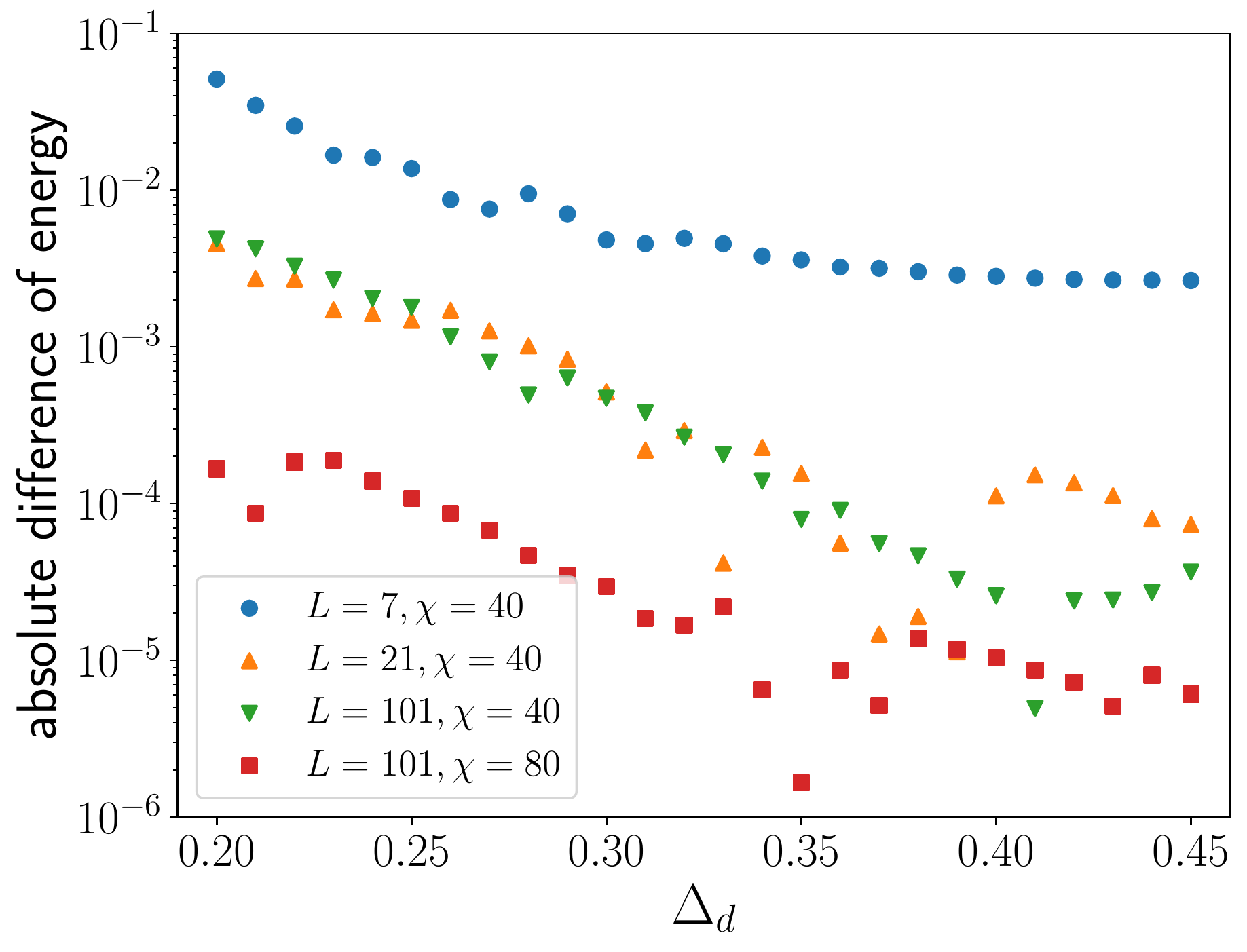}
\caption{Absolute difference of energy between those obtained from the correlation matrix of Gaussian fPEPS and direct tensor network contraction of an infinite fPEPS. The hole density is set to $\delta=0.16$ and the virtual fermion flavor number in the Gaussian fPEPS is $\Lambda=2$. $\chi$ is the bond dimension used in the variational uniform matrix product state algorithm~\cite{vumps}. This result reveals that for small system size, the difference mainly comes from the finite-size effect of Gaussian fPEPS and can be minimized by enlarging system size. If the system size is large enough, the difference comes from the finite bond dimension of environments in contraction. Nevertheless, the difference is quite small.
\label{fig:ET-EABD}}
\end{figure}
\begin{figure}[t]
\centering
        \includegraphics[width=\columnwidth]{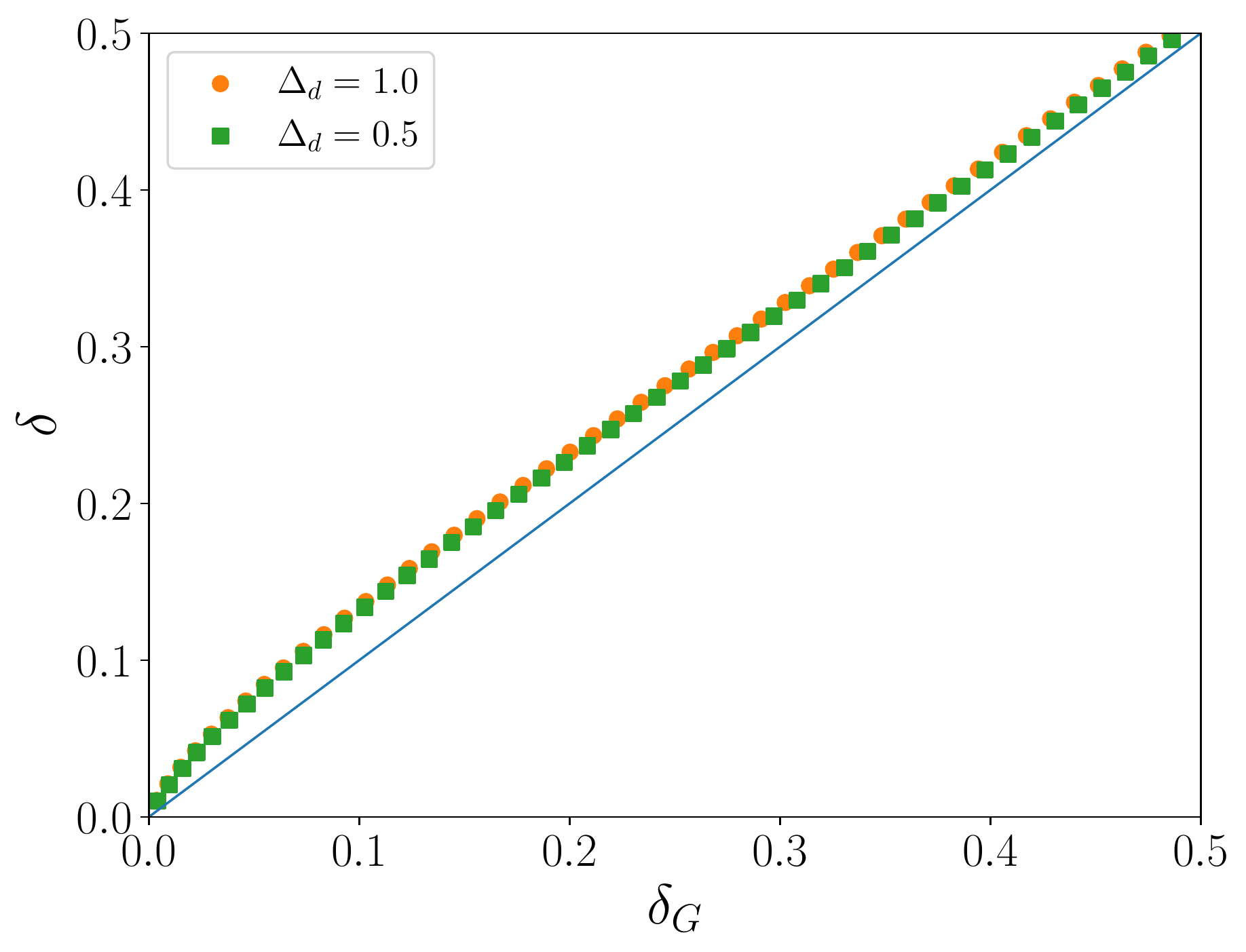}
\caption{Hole density before ($\delta$) and after ($\delta_{G}$) the Gutzwiller projection. Here, we consider $\Lambda=2$ and $L=101$. The deviation of points from the blue line represents the error of the Gutzwiller approximation for the hole fugacity in \Eq{eq:holefugacity}.
\label{fig:nafter_nbefore}}
\end{figure}

The BCS Hamiltonian in Eq.~\eqref{eq:HMF} has Dirac points at $\mathbf{k}=(\pm \tfrac{\pi}{2}, \pm \tfrac{\pi}{2})$. When optimizing $\Gamma$, we choose a suitable combination of the system size $L$ and boundary condition to avoid these exact zero modes, as we mentioned in Sec.~\ref{sec:construction_of_gfpeps}.

First, as a sanity check, the expectation value of the Hamiltonian in \Eq{eq:HMF} is calculated from optimized Gaussian fPEPS. The expectation value can be obtained in two different ways. The first way is through tensor network contraction, and the second way is using the correlation matrix [see \Eq{eq:expectation_HBCS}]. The result in Fig.~\ref{fig:ET-EABD} shows that the energies evaluated in these two ways agrees with each other. The small discrepancy is due to the finite bond dimension of environments in the tensor network contraction. Such discrepancy can be reduced systematically by enlarging the bond dimension kept in the contraction.

Next, we move on to the Gutzwiller projected states. We denote the density of holes of unprojected fPEPS as $\delta$ and that of Gutzwiller projected state as $\delta_G$, respectively.
With a suitable choice of hole fugacity $z$, we expect $\delta=\delta_G$.
Here, we examine the Gutzwiller approximated fugacity $z=2\delta/(1+\delta)$~\cite{Edegger2005}. The result in Fig.~\ref{fig:nafter_nbefore} shows that the approximated fugacity is smaller than needed since $\delta$ is larger than $\delta_{G}$. Thus, the fugacity should be enlarged to make sure that $\delta$ remains unchanged after the Gutzwiller projection.

The projected d-wave pairing state is believed to be a good candidate for the ground state of the $t$-$J$ model~\cite{Anderson2004} with the Hamiltonian~\cite{Zhang1988,Ogata2008}
\begin{equation}
H_{tJ}=-t \sum_{\langle \mathbf{i}, \mathbf{j}\rangle, \sigma} P_G \left(f_{\mathbf{i} \sigma}^{\dagger} f_{\mathbf{j} \sigma}+\mathrm { h.c. }\right) P_G+J \sum_{\langle \mathbf{i}, \mathbf{j}\rangle}\left(\vec{S}_{\mathbf{i}} \cdot \vec{S}_{\mathbf{j}}-\frac{n_{\mathbf{i}} n_{\mathbf{j}}}{4}\right),
\end{equation}
where $\vec{S_i}=\frac{1}{2}\sum_{ab}f^\dagger_{\mathbf{i},a} \boldsymbol{\sigma}_{ab}f_{\mathbf{i},b}$ and $n_{\mathbf{i}}= \sum_{\sigma}f^\dagger_{\mathbf{i}\sigma}f_{\mathbf{i}\sigma}$ are spin and charge density operators, respectively. We set $J/t=0.4$ ($t=1$) and carry out a variational calculation with the Gutzwiller projected d-wave ansatz.

Treating $\Delta_d$ in \Eq{eq:HMF} as the single variational parameter, we need to optimize the variational energy of $H_{tJ}$ (denoted as $E_{tJ}$) for each given hole concentration $\delta_{G}$. In the calculation, it is crucial to keep the same $\delta_{G}$ when varying $\Delta_d$ since the variational energy is very sensitive to $\delta_{G}$. We achieve this by tuning the fugacity using a bisection search as an inner loop for the variational optimization of $\Delta_d$.

\begin{figure}[t]
  {
  \begin{tikzpicture}

      \draw[cyan,line width=1.0pt] (-1,-0.5) -- (1,-0.5);
      \draw[cyan,line width=1.0pt] (-1,0.5) -- (1,0.5);

      \draw[green,line width=1.0pt] (-0.5,1) -- (-0.5,-1);
      \draw[green,line width=1.0pt] (0.5,1) -- (0.5,-1);

      \filldraw[fill=red!80!white,draw=red!80!white] (0.5,0.5) circle(0.16);
      \filldraw[fill=red!80!white,draw=red!80!white] (-0.5,0.5) circle(0.16);
      \filldraw[fill=red!80!white,draw=red!80!white] (0.5,-0.5) circle(0.16);
      \filldraw[fill=red!80!white,draw=red!80!white] (-0.5,-0.5) circle(0.16);
  \end{tikzpicture}
  }
\caption{The density of charge and paring strength of the Gutzwiller projected BCS state in the case of $\Lambda=2,\delta=0.16,\Delta_d=0.25$. The diameter of red circles scales with the quantity of density of holes. The width of cyan and green lines scales with the absolute value of paring strength where the sign of the paring strength is positive(negative) for the cyan(green) lines.}
\label{fig:state}
\end{figure}
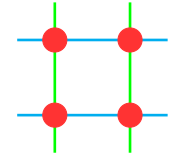

Figure~\ref{fig:state} shows the measured correlations for an optimized state.
The state is uniform and does not exhibit magnetic or charge order but hosts a d-wave superconducting order.

Directly observing a perfect dome shape in the order parameter $\frac{1}{\sqrt{2}}\left|\langle f_{\mathbf{i}\uparrow}f_{\mathbf{j}\downarrow} - f_{\mathbf{i}\downarrow}f_{\mathbf{j}\uparrow} \rangle\right|$ for neighboring sites $\mathbf{i},\mathbf{j}$ is difficult for some reasons. The convergence of the variational uniform matrix product state algorithm is difficult for small $\Delta_d$ and $\delta_G$. Thus, the range of investigated $\delta_G$ is limited to $(0.11,0.16)$. Besides, the tensor network contraction error makes it difficult to find optimal $\Delta_d$ and corresponding order parameter with high precision. As an outlook, it might be possible to improve the convergence and precision by exploiting the SU(2) symmetry of the fPEPS in tensor network contractions~\cite{u1gfpeps}.
\begin{figure}[t]
\centering
      \includegraphics[width=\columnwidth]{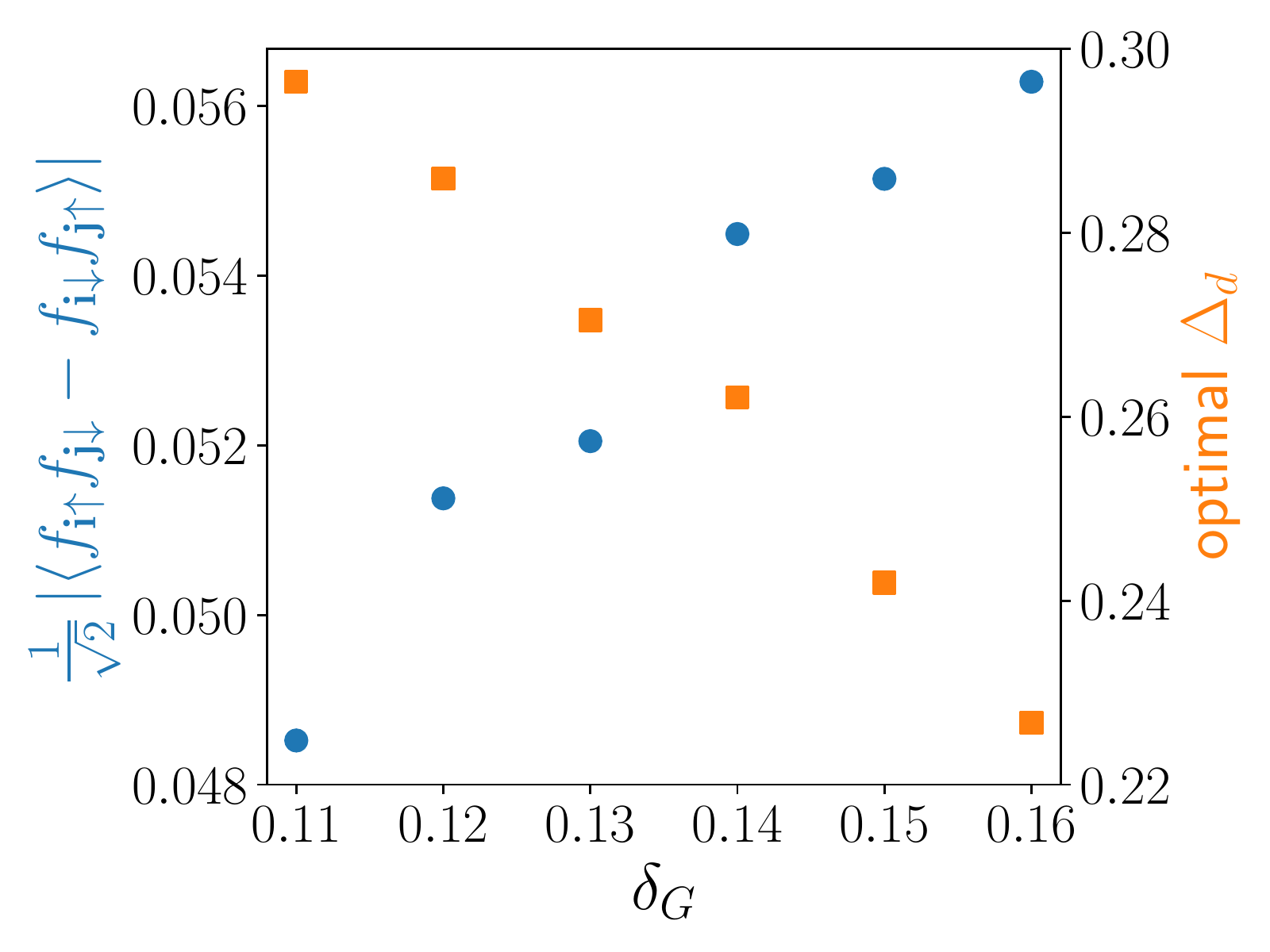}
\caption{Order parameter $\frac{1}{\sqrt{2}}\left|\langle f_{\mathbf{i}\uparrow}f_{\mathbf{j}\downarrow} - f_{\mathbf{i}\downarrow}f_{\mathbf{j}\uparrow} \rangle \right|$ (blue circle) for neighboring sites $\mathbf{i},\mathbf{j}$ and optimal $\Delta_d$ (orange square) versus $\delta_G$ for $\Lambda=2$ and $L=101$. The order parameter is obtained by using the tensor network contraction. For stability, the optimal $\Delta_d$ is determined via a second degree polynomial fitting near the optimal points.
\label{fig:OPandOptimalDelta}}
\end{figure}

\begin{figure}[t]
  \centering
  \resizebox{0.99\columnwidth}{!}{\includegraphics[width=\columnwidth]{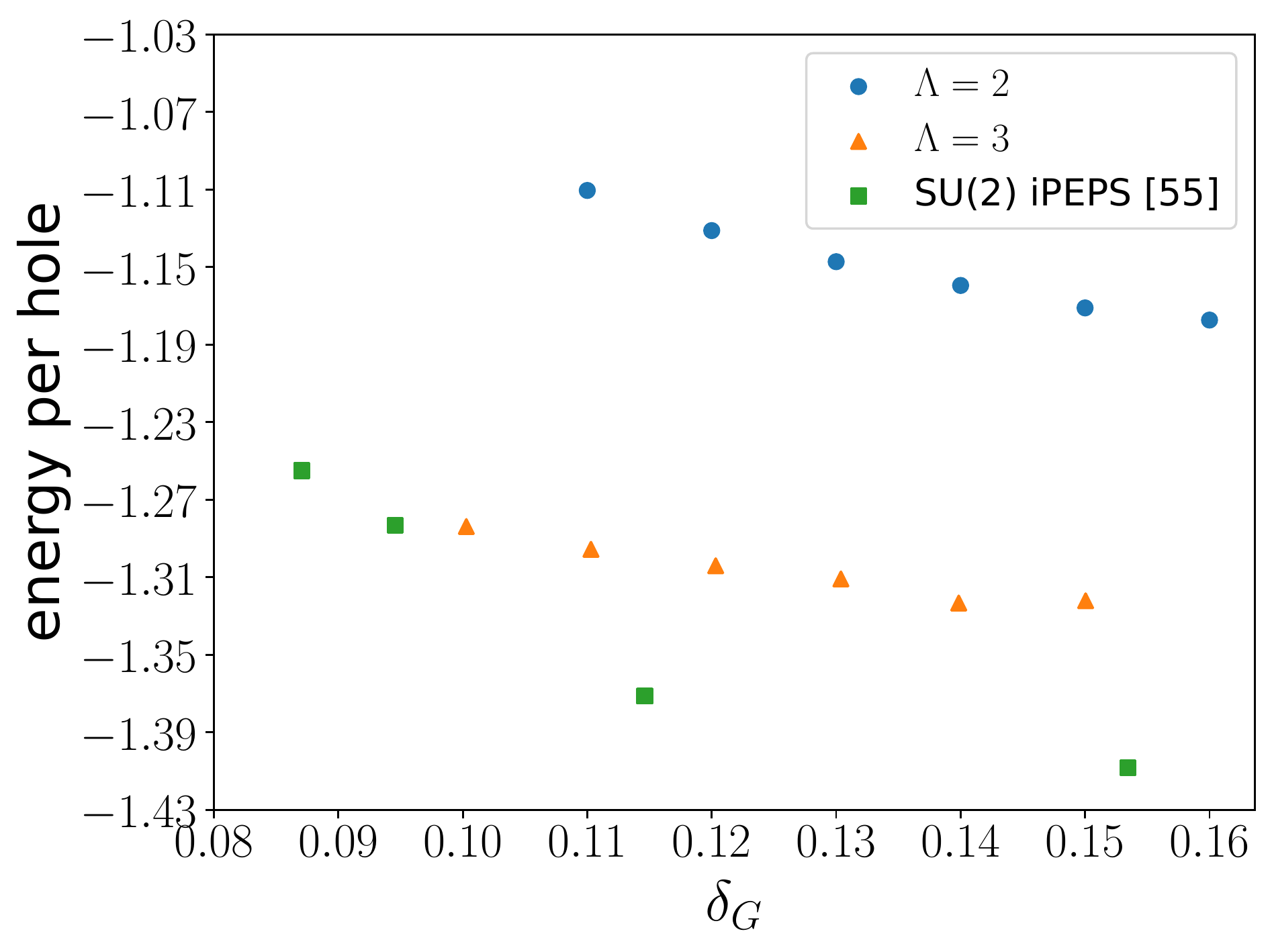}}
  \caption{Energy per hole. In $\Lambda=2$ cases, we vary $\Delta_d$ to minimize the energy of the $t$-$J$ model for a given density of holes $\delta_G$. Fugacity in the Gutzwiller projector is tuned for each $\Delta_d$ to ensure the final $\delta_G$ be the same, which is crucial in finding optimized $\Delta_d$. For $\Lambda=3$ cases, we used optimized $\Delta_d$ and solved fugacity in $\Lambda=2$ for each $\delta$ to avoid a costly optimization of $\Delta_d$ and solution of fugacity in $\Lambda=3$. Green squares are from SU(2) fPEPS calculations in~\cite{li2020ipeps}.
  \label{fig:energy}}
  \end{figure}

As the bisection search of $z$ is unaffordable for $\Lambda=3$, we investigate the order parameter for $\Lambda=2$. As shown in Fig.~\ref{fig:OPandOptimalDelta}, in the limited range of $\delta_G$, the optimal variational parameter $\Delta_d$ decreases with doping and the pairing order parameter increases with doping. On the other hand, the order parameter increases with $\Delta_d$ for a fixed doping. Putting these three facts together, it can be expected that for larger $\delta_G$, the order parameter will decrease as the optimal $\Delta_d$ decreases with doping and finally drop to zero. In conclusion, one would obtain a dome shape in the order parameter versus doping.

Lastly, we compute the energy per hole $(E_{tJ}+0.46778)/\delta_{G}$ for $\Lambda=2$ and $3$ where $-0.46778$ is the ground-state energy for the undoped case~\cite{PhysRevB.56.11678}. Increasing bond dimension results in a lower energy expectation value, reaching an energy per hole as low as $-1.32$. The energies are higher than those obtained from direct optimization of fPEPS in Refs.~\cite{Corboz2014, li2020ipeps}. Figure~\ref{fig:energy} shows the energy per hole as a function of doping. We compare the energy with the SU(2) fPEPS results of Ref.~\cite{li2020ipeps} obtained at $D^*=6$, which amounts to $D\approx 11$ without symmetry. Note that for $\Lambda=3$, our fPEPS obtained from the Gutzwiller ansatz has bond dimension $D=8$. Besides, our fPEPS tensor is uniform, while Ref.~\cite{li2020ipeps} uses a $2\times 2$ or $5\times 2$ unit cell. Constructing a tensor network representation of projected d-wave pairing states opens the way to provide good initial states in variational calculations besides offering diagnosis from the tensor network toolbox.\newline

\section{Summary and Discussions}
To summarize, we have developed a systematic method to construct the fPEPS representation of projected BCS states. Using the method, we investigated the physical properties of the projected d-wave pairing state on an infinite square lattice. In particular, the estimated variational energy for the $t$-$J$ model is comparable to the results from tensor network optimizations. The code implementation is available online~\footnote{\url{https://github.com/TensorBFS/Gaussian-fPEPS}}.

Our approach can also be applied to other classes of projected fermionic states, such as those obtained from Chern insulators and spin density wave states~\cite{PhysRevB.37.3774}. In certain cases, one would need an enlarged unit cell in the construction and contraction of tensor networks. In contrast to previous works~\cite{Verstraete2006a,Poilblanc2012,Schuch2012,Wang2013a,Poilblanc2014,Yang2015} which rely on an analytical PEPS representation of the RVB state, the present approach variationally solves a fermionic quadratic Hamiltonian of unprojected states using Gaussian fPEPS~\cite{Mortier2020}. Thus, the present approach has a broader range of applications and can be systematically improved by increasing the bond dimension of the Gaussian fPEPS tensor.

Partial Gutzwiller projection which suppresses but does not eliminate double occupancies~\cite{PhysRevLett.10.159} can also be straightforwardly implemented. These states are relevant to the study of Gossamer superconductors~\cite{Laughlin2006,PhysRevLett.90.207002} and Hubbard models. In principle, long-range Jastrow factors beyond the onsite Gutzwiller projection can be applied by using its tensor network form~\cite{Li2019m}. This class of states is considered to be crucial for describing Mott transitions~\cite{Capello2005}. Further investigations of these states with our method are left for future works.

\section{Acknowledgments}
We thank Jan von Delft, Jheng-Wei Li, and Yi Zhou for useful discussions. This project is supported by National Key Projects for Research and Development of China Grants No.~2021YFA1400400 and No.~2022YFA1403900, the Strategic Priority Research Program of the Chinese Academy of Sciences under Grant No.~XDB30000000, the National Natural Science Foundation of China under Grant No.~T2121001, the Deutsche Forschungsgemeinschaft (DFG) through project A06 of SFB 1143 (Project No.~247310070), and the Youth Innovation Promotion Association CAS under Grant No.~2021004.

\bibliography{refs}

\appendix

\section{Proof of \Eq{eq:ABD}}\label{appendix:proof-abd}
To establish the relationship between the correlation matrix of the Gaussian fPEPS and that of the fiducial state in \Eq{eq:LocalGamma}, we prove a general result below.

Let us consider a fermionic Gaussian state $|\omega\rangle$ living in the Hilbert space $\mathcal{H}_2$ with virtual Majorana modes $d_l$ ($l=1,\ldots,2m$) and another fermionic Gaussian state $|A\rangle$ living in the composite Hilbert space $\mathcal{H}_1\otimes\mathcal{H}_2$ including both physical and virtual Majorana modes~\footnote{Rigorously speaking, the composite Hilbert space $\mathcal{H}_1\otimes\mathcal{H}_2$ of fermions should be viewed as a $\mathbb{Z}_2$-graded tensor product rather than a usual tensor product. However, this distinction is unnecessary for our purpose here, as the operators under consideration are density operators involving even number of Majorana modes.}, where the physical modes are denoted as $c_j$ ($j=1,\ldots,2n$).

The overlap $\langle\omega|A\rangle$ is also a fermionic Gaussian state and lives in the physical Hilbert space $\mathcal{H}_1$. To calculate its correlation matrix, it is convenient to work with density operators and write the density operator of $\langle\omega|A\rangle$ as
\begin{equation}
\rho_f(c) = \langle\omega|A\rangle \langle A|\omega\rangle
    = \mathrm{tr}_{\mathcal{H}_2} [\rho_{A}(c,d) \rho_{\omega}(d)],
\label{eq:partial-trace}
\end{equation}
where $\mathrm{tr}_{\mathcal{H}_2}$ is the partial trace over $\mathcal{H}_2$ and $\rho_{\omega}$ ($\rho_{A}$) is the density operator of $|\omega\rangle$ ($|A\rangle$).

We shall establish a Grassmann integration approach to calculate the partial trace in Eq.~\eqref{eq:partial-trace}. This approach uses the Grassmann representations of the Gaussian density operators $\rho_{\omega}$ and $\rho_{A}$~\cite{Bravyi2005},
\begin{equation}
g_{\omega}(\tau)=\frac{1}{2^m}\exp \left(\frac{i}{2}\tau^T\Gamma_{\omega}\tau\right),
\label{eq:g_omega}
\end{equation}
and
\begin{equation}
g_{A}(\theta,\zeta)=\frac{1}{2^{n+m}}\exp \left[\frac{i}{2}\begin{pmatrix}
\theta^T & \zeta^T
\end{pmatrix}
\begin{pmatrix}
A & B \\
-B^T & D
\end{pmatrix}
\begin{pmatrix}
\theta \\
\zeta
\end{pmatrix}
\right],
\label{eq:g_A}
\end{equation}
where $\theta = (\theta_1,\ldots,\theta_{2n})^T$ is a vector of real Grassmann variables for the physical modes $c = (c_1,\ldots,c_{2n})^T$ and, similarly, $\tau$ and $\zeta$ include Grassmann variables for virtual modes. Here $\Gamma_{\omega}$ and $\begin{pmatrix}
A & B \\
-B^T & D
\end{pmatrix}$ are the correlation matrices of $\rho_{\omega}$ and $\rho_{A}$, respectively.

In the Grassmann representation, the partial trace over $\mathcal{H}_2$ in Eq.~\eqref{eq:partial-trace} can be calculated using a Grassmann integration as follows:
\begin{equation}
g_f\left(\theta\right) =  (-2)^m \int D \zeta D \tau \; e^{\zeta^{T} \tau} g_A(\theta, \zeta) g_\omega\left(\tau \right),
\label{eq:ggg}
\end{equation}
where $g_f(\theta)$ is the Grassmann representation of the (unnormalized) density operator for $\langle\omega|A\rangle$ and the integration is over Grassmann variables associated with virtual modes, with the notation $\int D \zeta D \tau = \int \mathrm{d}\zeta_{2m}\cdots \mathrm{d}\zeta_{1} \mathrm{d}\tau_{2m} \cdots \mathrm{d}\tau_{1}$. The proof of \eqref{eq:ggg} is done by comparing the outcomes of Eqs.~\eqref{eq:partial-trace} and \eqref{eq:ggg}, where the former can be computed by expanding $\rho_{A}$ and $\rho_{\omega}$ in terms of Majorana operators and performing the partial trace, and the latter is computed by expanding the Grassmann exponential and carrying out the integration over Grassmann variables one by one.

After substituting Eqs.~\eqref{eq:g_omega} and \eqref{eq:g_A} into Eq.~\eqref{eq:ggg} and performing the Gaussian integration over $\zeta$ and $\tau$, we obtain
\begin{equation}
g_f\left(\theta\right) = \frac{1}{2^{n+m}} \operatorname{Pf}\left(\Gamma_{\omega}\right) \operatorname{Pf}\left(D-\Gamma_{\omega}^{-1}\right) \exp \left(\frac{i}{2} \theta^{T} \Gamma_{\mathrm{f}} \theta\right),
\end{equation}
where $\mathrm{Pf}$ denotes the Pfaffian for antisymmetric matrices and $\Gamma_{\mathrm{f}}$ is the correlation matrix of $\langle\omega|A\rangle$ and has the following explicit form:
\begin{align}
\Gamma_{\mathrm{f}} &=A+B\left(D-\Gamma_{\omega}^{-1}\right)^{-1} B^{T} \nonumber \\
&= A+B\left(D+\Gamma_{\omega}\right)^{-1} B^{T}.
\label{eq:g_f}
\end{align}
Here we used $\Gamma_{\omega}^{-1} = -\Gamma_{\omega}$ as $|\omega\rangle$ is a pure state.

We note that \Eq{eq:ABD} is just the Fourier transformed version of \eqref{eq:g_f} taking into account the translation invariance. It is also worth mentioning that Eq.~\eqref{eq:g_f} agrees with Eq.~(C8) in Ref.~\cite{Hackenbroich2020} but differs from those in Refs.~\cite{Kraus2010,Wahl2013,Wahl2014} by the sign in front of $\Gamma_{\omega}$. This is due to that the Gaussian fPEPS projector in Refs.~\cite{Kraus2010,Wahl2013,Wahl2014} is defined through the Gaussian map~\cite{Bravyi2005}, which is different from Eq.~\eqref{eq:fpeps}.

\section{Overlap-based translation}\label{appendix:overlap-translation}
Instead of diagonalizing the fiducial Hamiltonian in the Fock space, we provide below an alternative approach to obtain the explicit tensor form of $\ket{A_{\mathbf{i}}}$. This approach is based on an overlap formula between fermionic Gaussian states and Fock states.

To start with, we rewrite the fiducial Hamiltonian~\eqref{eq:fiducial-Hamiltonian} in a Bogoliubov-de Gennes (BdG) form by converting the Majorana operators $\gamma$ back to original complex fermions
\begin{equation}
H = \left(\begin{array}{ll}
c^{\dagger} & c
            \end{array}\right) \mathcal{H}
\left(\begin{array}{ll}
c \\
c^\dag
            \end{array}\right),
\label{eq:BdG}
\end{equation}
where $(c^{\dagger} \; c)$ is a row vector of fermionic creation and annihilation operators for $4\Lambda+2$ physical and virtual fermions at a local site (see Sec.~\ref{sec:construction_of_gfpeps}). The BdG single-particle Hamiltonian $\mathcal{H}$ is diagonalized by a Bogoliubov transformation
\begin{equation}
\begin{aligned}
\left(\begin{array}{ll}
a^{\dagger} & a
            \end{array}\right)=\left(\begin{array}{ll}
c^{\dagger} & c
            \end{array}\right)\left(\begin{array}{ll}
U & V^{*} \\
V & U^{*}
\end{array}\right),
\end{aligned}
\label{eq:A1}
\end{equation}
where $(a^{\dagger} \; a)$ includes creation and annihilation operators of Bogoliubov modes and the Bogoliubov matrices $U$ and $V$ satisfy
\begin{equation}
U^\dagger U+V^\dagger V = \mathbb{I}, \quad U^TV+V^TU = \mathbb{O}
\end{equation}
with $\mathbb{I}$ ($\mathbb{O}$) being the identity (zero) matrix.

The local fiducial state $\ket{A_{\mathbf{i}}}$, being the ground state of $H$, is annihilated by all $a$-fermion annihilation operators. According to Eq.~\eqref{eq:LocalState}, the tensor form of $\ket{A_\mathbf{i}}$ can be obtained by calculating each coefficient $A^f_{uldr}$ which is the overlap $\bra{f}\otimes\braket{uldr|A_\mathbf{i}}$. $\bra{f}\otimes\bra{uldr}$ denotes a Fock state in the $c$-fermion basis and can generally be written as ${_{c}\bra{0}} c_{i_
{M'}} \cdots c_{i_1}$ with $0\leq M' \leq 4\Lambda+2$ and $1\leq i_1 < \cdots < i_{M'} \leq 4\Lambda+2$, where ${_{c}\bra{0}}$ is the $c$-fermion vacuum.

Such overlap calculation can be carried out by rewriting the Bogoliubov vacuum $\ket{A_{\mathbf{i}}}$ in a suitable form~\cite{Bertsch_Robledo_2011}, with the help of the Bloch-Messiah decomposition~\cite{bloch-messiah}
\begin{equation}
\begin{aligned}
&\left(\begin{array}{cc}
U& V^{*} \\
V & U^{*}
\end{array}\right)=\left(\begin{array}{cc}
D & 0 \\
0 & D^{*}
\end{array}\right)\left(\begin{array}{ll}
\bar{U} & \bar{V} \\
\bar{V} & \bar{U}
\end{array}\right)\left(\begin{array}{cc}
C & 0 \\
0 & C^{*}
\end{array}\right),
\end{aligned}
\label{eq:bloch_messiah}
\end{equation}
where $D$ and $C$ are unitary matrices and $\bar{U}$ and $\bar{V}$ are real matrices with the following form:
\begin{equation}
\bar{U} = \begin{pmatrix}
\mathbb{O} & & \\
& \bigoplus_p u_p\sigma^0 & \\
& & \mathbb{I}
\end{pmatrix}, \; \bar{V} = \begin{pmatrix}
\mathbb{I} & & \\
& \bigoplus_p iv_p\sigma^y & \\
& & \mathbb{O}
\end{pmatrix}.
\end{equation}
Here $u_p$ and $v_p$ are positive and satisfy $u_p^2 + v_p^2 = 1$, $\mathbb{I}$ ($\mathbb{O}$) is the identity (zero) matrix, and $\sigma^0$ and $\sigma^y$ are the $2\times 2$ identity and Pauli matrices. For our purpose, it is convenient to truncate the $\mathbb{I}$ ($\mathbb{O}$) block in $\bar{U}$ ($\bar{V}$) to obtain matrices
\begin{equation}
\bar{U}' = \begin{pmatrix}
\mathbb{O} &  \\
& \bigoplus_p u_p\sigma^0 \\
\end{pmatrix}_{M\times M}, \;
\bar{V}' = \begin{pmatrix}
\mathbb{I} & \\
& \bigoplus_p iv_p\sigma^y \\
\end{pmatrix}_{M\times M}.
\end{equation}
Following Ref.~\cite{Jin2021d}, $\ket{A_\mathbf{i}}$ is represented as
\begin{equation}
\ket{A_\mathbf{i}} = \frac{1}{\prod_p v_p} b_1 \cdots b_M \ket{0}_c ,
\label{eq:bdg_state_formalism}
\end{equation}
where $b = c^\dagger D' \bar{V}'+c(D')^* \bar{U}'$. Here $D'$ is a $(4\Lambda+2)\times M$ matrix obtained by keeping the first $M$ columns of the unitary matrix $D$ [see Eq.~\eqref{eq:bloch_messiah}], which is an isometry satisfying $(D')^\dagger D' = \mathbb{I}_{M \times M}$.

With these results in hand, the tensor entry $A^f_{uldr}$ is calculated with Wick's theorem as follows:
\begin{align}
\bra{f}\otimes\braket{uldr|A_\mathbf{i}} &= \frac{1}{\prod_p v_p} \, {_{c}\bra{0}} c_{i_
{M'}} \cdots c_{i_1} b_1 \cdots b_M \ket{0}_c \nonumber \\
&= \frac{(-1)^{\frac{1}{2}M'(M'-1)}}{\prod_{p}v_{p}} \, {_{c}\bra{0}} c_{i_1} \cdots c_{i_
{M'}} b_1 \cdots b_M \ket{0}_c \nonumber \\
&= \frac{(-1)^{\frac{1}{2}M'(M'-1)}}{\prod_{p}v_{p}}\mathrm{Pf}\begin{pmatrix}
\mathbb{O}_{M'\times M'} & \mathbb{R}_{M'\times M} \\
-\mathbb{R}^T_{M\times M'} & \mathbb{Q}_{M\times M}
\end{pmatrix},
\label{eq:Pfaffian_formula}
\end{align}
where $\mathbb{R}_{m'm}={_{c}\bra{0}} c_{i_{m'}}b_m\ket{0}_c=(D'\bar{V}')_{i_{m'}m}$ and $\mathbb{Q}_{m'm}={_{c}\bra{0}} b_{m'} b_m\ket{0}_c=(\bar{U}'\bar{V}')_{m'm}$. Note, however, that the overlap in Eq.~\eqref{eq:Pfaffian_formula} vanishes if $M$ and $M'$ have different parity (this automatically encodes the fermion parity symmetry). The Pfaffian formula~\eqref{eq:Pfaffian_formula} allows us to calculate the explicit tensor form of $\ket{A_{\mathbf{i}}}$. In our program, the Pfaffian is calculated with an algorithm provided by M.~Wimmer~\cite{10.1145/2331130.2331138}.

We can compare the pros and cons of Hamiltonian-based translation and overlap-based translation: The complexity of translation is quantified by $N=4\Lambda+2$. When employing an overlap-based algorithm, we need to calculate matrix Pfaffians for $2^{N-1}$ times. As the size of each matrix is of order $2N \times 2N$, the computational cost of the overlap-based translation is $O(N^3\times 2^{N-1})$. This complexity can be further reduced to $O(N^2\times 2^{N-1})$ if a fast update technique for the Pfaffian is used (see, e.g., Ref.~\cite{morita2015}). Meanwhile, in the Hamiltonian-based algorithm, the computational cost of obtaining the dominant eigenvector of a $2^N\times 2^N$ matrix scales as $O((2^N)^{2})=O(2^{2N})$.

When $\Lambda$ is large, the Hamiltonian-based algorithm in Sec.~\ref{sec:translation} has a larger computational cost, but it is nevertheless more straightforward and intuitive. If $\Lambda$ is so large that an exact diagonalization of the fiducial Hamiltonian is not feasible, one might also employ the density matrix renormalization group method~\cite{White1992} to obtain an approximation of $\ket{A_{\mathbf{i}}}$.

\end{document}